\input{aipcheck}

\documentclass[
    ,final            
  ]
  {aipproc}

\layoutstyle{6x9}

\usepackage{bm}
\usepackage{amsmath}

\begin{document}

\title{Quasiparticles and $Z(N)-$ lines in Hot Yang-Mills theories}


\author{Marco Ruggieri}{
  address={Department of Physics and Astronomy, University of Catania, Via S. Sofia 64, I-95125 Catania}
}

\author{Paolo Alba}{
  address={Department of Physics and Astronomy, University of Catania, Via S. Sofia 64, I-95125 Catania}
}

\author{Paolo Castorina}{
  address={Department of Physics and Astronomy, University of Catania, Via S. Sofia 64, I-95125 Catania}
  ,altaddress={INFN, Sezione di Catania, I-95123 Catania, Via Santa Sofia 64, Italy} 
}

\author{Salvatore Plumari}{
  address={Department of Physics and Astronomy, University of Catania, Via S. Sofia 64, I-95125 Catania}
  ,altaddress={INFN-Laboratori Nazionali del Sud, Via S. Sofia 62, I-95123 Catania} 
}

\author{Claudia Ratti}{
  address={Dipartimento di Fisica, Universita` degli Studi
di Torino e INFN, Sezione di Torino, via Giuria 1, I-10125 Torino}
}

\author{Vincenzo Greco}{
  address={Department of Physics and Astronomy, University of Catania, Via S. Sofia 64, I-95125 Catania}
  ,altaddress={INFN-Laboratori Nazionali del Sud, Via S. Sofia 62, I-95123 Catania} 
}

\begin{abstract}
In this talk we review, the quasiparticle description of the hot Yang-Mills theories,
in which the quasiparticles propagate in (and interact with) a background field
related to $Z(N)$-lines.
We compare the present description with a more common one in which the 
effects of the $Z(N)$-lines are neglected. We show that it is possible to take
into account the nonperturbative effects at the confinement transition temperature
even without a divergent quasiparticle mass.
\end{abstract}

\maketitle

\section{Introduction}
In recent years, the interest in understanding the
thermodynamical properties of strong nonabelian mediums has noticeably 
increased.
Lattice simulations of $SU(3)$ Yang-Mills 
theories~\cite{Datta:2010sq,Boyd:1996bx,Boyd:1995zg,Borsanyi:2011zm,Borsanyi:2012ve}
agree about the onset of the deconfinenemt phase transition
at $T= T_c \approx 270$ MeV. Below $T_c$, lattice data suggest the thermodynamics of the 
Yang-Mills theory is dominated by the lowest 
lying glueballs superimposed to a Hagedorn spectrum~\cite{Borsanyi:2012ve}. 
For what concerns the high temperature phase, the picture
is not so clear. As a matter of fact, the perturbative regime
is realized for very large temperatures,
signalling the nonperturbative
nature of the gluon medium above the critical temperature \cite{zwanziger2}.
This makes the identification
of the correct degrees of freedom of the gluon plasma, in proximity
as well as well beyond the critical temperature, a very complicated task.

Beside resummation schemes based on the Hard Thermal Loop (HTL) 
approach~\cite{HTL1,HTL3,HTL4,HTL7,Andersen_HTLpt,Andersen:2010ct,Andersen:2009tc}
and on the effective potential approach~\cite{Braun:2007bx},
the quasiparticle approach to the thermodynamics of QCD has attracted a discreet 
interest recently, see for example Refs.~\cite{Gorenstein:1995vm,PKPS96, Levai:1997yx,bluhm-qpm,
Meisinger:2003id,Castorina:2011ra,Giacosa:2010vz,Peshier:2005pp,redlich,
Plumari:2011mk,Filinov:2012pt,Castorina:2007qv,Castorina:2011ja,Brau:2009mp,
Cao:2012qa,Bannur:2006hp,Sasaki:2012bi,Ruggieri:2012ny} and
references therein. In such an approach, one identifies the degrees of 
freedom of the deconfinement phase with transverse gluons;
the strong interaction in such a nonperturbative regime
is taken into account through an effective temperature-dependent mass for the gluons themselves.
Generally speaking, it is not possible to compute the
self-energy of gluons exactly in the range of temperature
of interest; as an obvious consequence, the location of the poles
of the propagator in the complex momentum plane, hence (roughly
speaking) the gluon mass, is an unknown function. For this reason,
one usually assumes an analytic dependence of the gluon mass on the
temperature, leaving few free parameters which are then fixed
by fitting the thermodynamical data of Lattice simulations. 

The advantage of such an approach is, at least, twofolds. 
Firstly, it is of a theoretical interest in itself to understand
which are the effective degrees of freedom of the Yang-Mills 
theory at finite temperature and in particular the evolution of
non perturbative effects with temperature.
Furthermore, once a microscopic description of the plasma is identified, 
it is also possible to include the implied dynamics into a transport theory capable of directly 
simulating the expanding fireball produced in heavy ion collisions
computing the collective properties, as well as the chemical 
composition of the fireball as a function of time~\cite{Scardina:2012hy}. 

In this talk, we discuss the quasiparticle picture of the finite
temperature gluon medium, by adding the interaction of gluons 
with a background Polyakov loop, or more generally speaking, with a background
of traced $Z(N)-$lines according to the nomenclature given in~\cite{Pisarski:2006hz}. 
Within our picture, the common view of the deconfinement phase of $SU(3)$ theory as
a gluon plasma, in which gluons are the relevant degrees of freedom, 
is replaced by a new one in which gluons propagate in a background of Polyakov loops,
the latter being important degrees of freedom as much as the gluons are, since they
embed nonperturbative informations about the 
deconfinement phase transition~\cite{Polyakov:1978vu,Susskind:1979up,Svetitsky:1982gs,Svetitsky:1985ye}.
This should be compared with the standard quasiparticle (plasma-like) picture, in which all the dynamics 
is taken into account by means of temperature dependent masses \cite{PKPS96,Levai:1997yx, bluhm-qpm,redlich,Castorina:2011ra,
Castorina:2011ja, Plumari:2011mk}, leading to diverging
or steadily increasing masses
as $T\rightarrow T_c$.
We show that combining a $T$-dependent quasiparticle mass, $M(T)$, with the Polyakov
loop dynamics results in a quite different behavior of $M(T)$ as $T\rightarrow T_c$.

\section{The model}
The system we consider~\cite{Ruggieri:2012ny} consists of a gas of gluon quasiparticles,
propagating in a background of Polyakov loops. For our purposes, considering fundamental and adjoint
loops is enough. The free energy of the model
is expressed as a linear combination of two contributions:
the first one describes the thermodynamics of the Polyakov loop; the second one,
on the other hand, is the contribution of gluon quasiparticles coupled
to the Polyakov loop. We specify the two terms of the effective potential
below. 

For what concerns the Polyakov loop effective potential, following~\cite{Abuki:2009dt,Zhang:2010kn}
we consider the action of a pure matrix model~\cite{Gupta:2007ax}, 
\begin{equation}
S_{PM}[L] = -N_c^2 e^{-a/T}\bm\sum_{\bm x,\bm y} \ell_F(\bm x)\ell_F^*(\bm x + \bm y)~,
\label{eq:sc1}
\end{equation}
where $\bm x,\bm y$ denote the lattice site and its nearest neighbors respectively. Moreover
$\ell_F(\bm x) = \text{Tr}L_F(\bm x)/N_c$
corresponds to the traced Polyakov loop in the fundamental representation. The Polyakov line
in the representation ${\cal R}$ of the gauge group is given by
\begin{equation}
L_{\cal R}(\bm x) = {\cal P}\exp\left[i\int_0^{1/T}d\tau~A_4^a(\bm x,\tau)T_{a,{\cal R}}\right]~,
\label{eq:RRR}
\end{equation}
where $T_{a,{\cal R}}$ are the gauge group generators in the representation ${\cal R}$, and $A_4^a$
corresponds to a background euclidean gluon field.

The model specified by Eq.~\eqref{eq:sc1} corresponds to a matrix model, the untraced Polyakov loops 
$L(\bm x)$ corresponding to the dynamical degrees of freedom. In this study we limit ourselves to the
one-loop approximation; at this level, the effective potential for the Polyakov loop reads~\cite{Ruggieri:2012ny}
\begin{eqnarray}
\Omega_{PM} &=& b T\left(-6N_c^2 e^{-a/T}\langle\ell_F\rangle^2
+N_c\left[\alpha \langle\ell_F\rangle \right] 
-\log\int dL~e^{N_c\left[\alpha \Re \ell_F \right]}\right)~.
\label{eq:MFWreal}
\end{eqnarray}
Here $\alpha$ corresponds to a variational parameter, whose value at a given temperature
is determined by the condition $\partial\Omega/\partial\alpha=0$; the average $\langle\ell_F\rangle$ is
a function of $\alpha$, which we compute numerically. Besides, $a$ and $b$ in the above equation
are treated as free parameters, whose value will be specified below. 

The transverse gluon quasiparticle contribution to the thermodynamic potential 
reads~\cite{Meisinger:2003id,Sasaki:2012bi,Ruggieri:2012ny}
\begin{equation}
\Omega_{qp} = 2T\int\frac{d^3k}{(2\pi)^3}\text{Tr}_A\log\left(1-L_A~e^{-E(k)/T}\right)~.
\label{eq:gqp}
\end{equation}
In the above equation, $L_A$ corresponds to the Polyakov line in the adjoint representation,
as defined in Eq.~\eqref{eq:RRR}, and the trace is taken over the indices of the adjoint representation
of the gauge group. Moreover, the quasiparticle energy is given by $E(k) = \sqrt{k^2 + M^2}$,
where $M$ is supposed to arise from non-perturbative medium effects. 
In the case of very high temperature, where the gauge theory is in the perturbative regime and thus $M = 0$,
it can be proved that Eq.~\eqref{eq:gqp} is the effective potential for the adjoint 
Polyakov loop~\cite{Weiss:1980rj}. At lower temperatures, where non-perturbative effects are important,
Eq.~\eqref{eq:gqp} must be postulated as a starting point for a phenomenological description of the 
thermodynamics of the gluon plasma. In this article, we make use of a temperature depentent mass whose analytic form
is given by~\cite{Peshier:2005pp,Ruggieri:2012ny}
\begin{equation}
M^2 = \frac{8\pi^2 T^2}{11\log\left[\lambda(T-w)\right]^2}~.
\label{eq:mass}
\end{equation}
The parameters $w,\lambda$ are determined by requiring that, for pressure and energy density, the mean quadratic deviation 
between our theoretical computation and the Lattice data of~\cite{Boyd:1996bx} is a minimum.

In order to have a combined description of the gluon plasma around the critical temperature,
in terms of the Polyakov loop (which acts as a background field) and of gluons quasiparticles
(which propagate in the Polyakov loop background), we add Eq.~\eqref{eq:gqp} to
the strong coupling inspired potential in Eq.~\eqref{eq:MFWreal}. In order to do this,
and to be consistent at the same time with the Weiss mean field procedure outlined above,
we follow the lines of~\cite{Abuki:2009dt} and write the thermodynamic potential as
\begin{equation}
\Omega = \Omega_{PM} + \langle\Omega_{qp}\rangle~,
\label{eq:FULL1}
\end{equation} 
where the averaged $\Omega_{qp}$ has to be understood as
\begin{equation}
\langle\Omega_{qp}\rangle = 2T\int\frac{d^3k}{(2\pi)^3}~
\log\left\langle\text{det}_A\left(1-L_A~e^{-E/T}\right)\right\rangle~.
\label{eq:SU3qp}
\end{equation}
We have checked numerically that in the deconfinement phase,
at very large temperature $\langle\ell_A\rangle \approx 1$; as a consequence, the thermal distribution of the quasigluons
approaches that of a perfect gas of massive particles in this limit. 
However, the thermodynamics of the system in the full range of temperature
considered here remains different from the one of a mere massive gas because of the
Polyakov background mean field.

\section{Results}
\begin{figure}[t!]
\includegraphics[width=7.0cm]{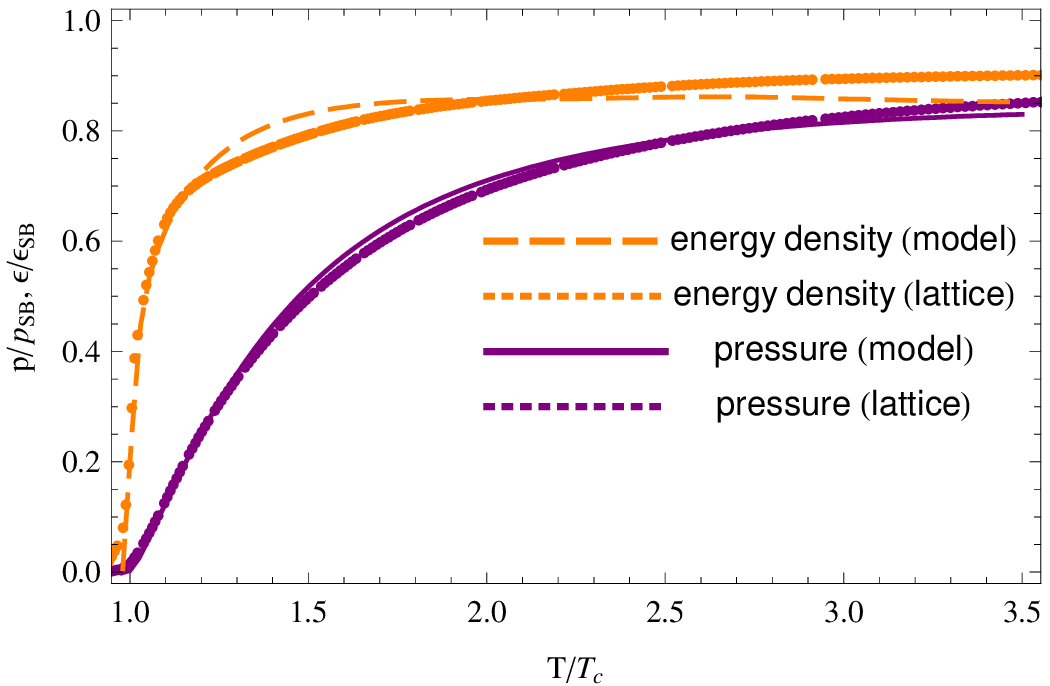}~~
\includegraphics[width=7.0cm]{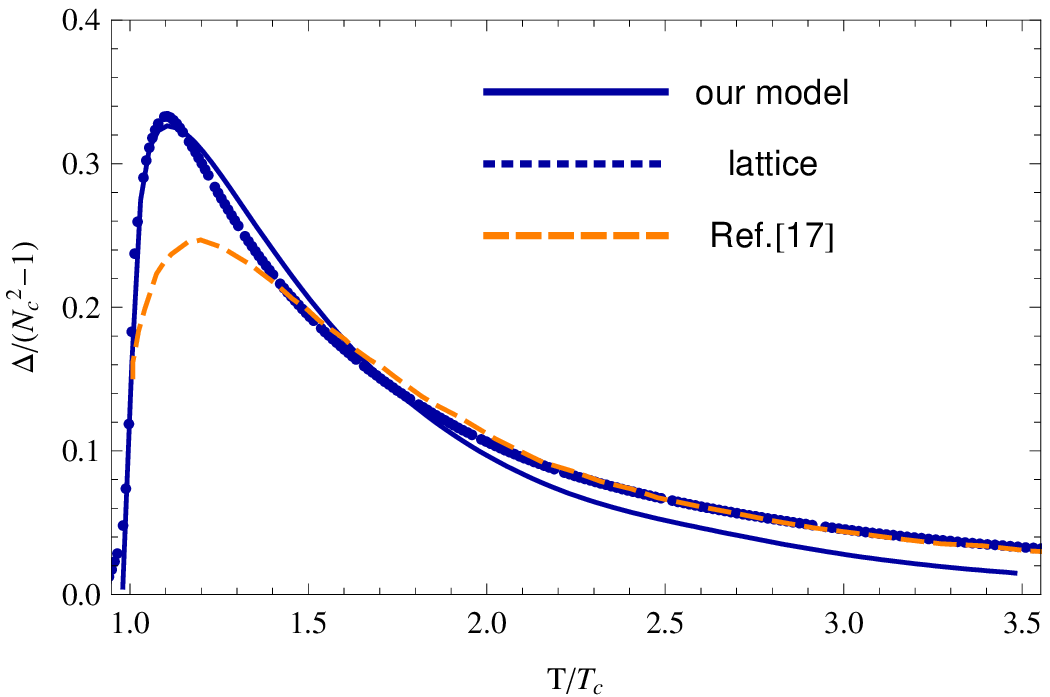}
\caption{\label{Fig:compSU3} {\em Left panel.} Pressure (indigo, solid line) and energy density (orange, dashed line), 
normalized to the 
respective Stefan-Boltzmann values, as a function of temperature (measured in units of the critical temperature $T_c = 270$ MeV),
for the SU(3) case. Dots correspond to lattice data taken from Ref.~\cite{Boyd:1996bx};
solid lines are the result of our numerical computation within the quasiparticle model.
{\em Right panel.} Interaction measure per degree of freedom as a function of temperature.
The dashed orange line corresponds to the result of Ref.~\cite{Meisinger:2003id}.
Figure adapted from Ref.~\cite{Ruggieri:2012ny}.}
\end{figure}

In this talk, we focus on our results on thermodynamical quantities (pressure, energy density and
interaction measure) and on the gluon quasiparticle mass; we refer to~\cite{Ruggieri:2012ny}
for more details. In the left panel of Fig.~\ref{Fig:compSU3} we plot the pressure $p=-\Omega$ (indigo data) and energy density (orange data), normalized to the 
Stefan-Boltzmann value, as a function of temperature (measured in units of the critical temperature),
for the SU(3) case. Given the pressure, the energy density is computed
by virtue of the thermodynamical relation performing the pertinent total derivative of the temperature, $\varepsilon = -p + T dp/dT$.
Of the four parameters in our model, one of them is fixed in order to reproduce the
first order phase transition at $T=270$ MeV; the remaining three parameters are fixed 
in order to require that the mean quadratic deviation for pressure, energy density and interaction
measure between our computation and the Lattice data is minimized.
This procedure leads to the numerical values $a=901.9$ MeV, $b=(157.5$ MeV$)^3$, $w=240$ MeV
and finally $\lambda = 0.25$ MeV$^{-1}$.  These parameters produce a gluon mass $M=360 \,$MeV at $T = T_c$. 
In the figure, $p_{SB}=(N_c^2 - 1) \pi^2 T^4/45$ and $\epsilon_{SB}=3 p_{SB}$.
Dots correspond to lattice data taken from Ref.~\cite{Boyd:1996bx};
solid lines are the result of our numerical computation within the quasiparticle model.

In the right panel of Fig.~\ref{Fig:compSU3} we compare our results for the interaction
measure, $\Delta = (\varepsilon - 3p)/T^4$, with the lattice data (represented by
dots in the figure). We notice that our model reproduces fairly well the peak of the
interaction measure observed on the lattice in the critical region. We also compare
our results with the ones obtained in~\cite{Meisinger:2003id}, where 
a phenomenological potential for the Polyakov loop is considered instead of our
matrix model, and the usual mean field approximation is used. 
The comparison shows that 
the peak of the interaction measure, where nonperturbative effects are important, 
is much better reproduced in our case.

\begin{figure}[t!]
\includegraphics[width=7.5cm]{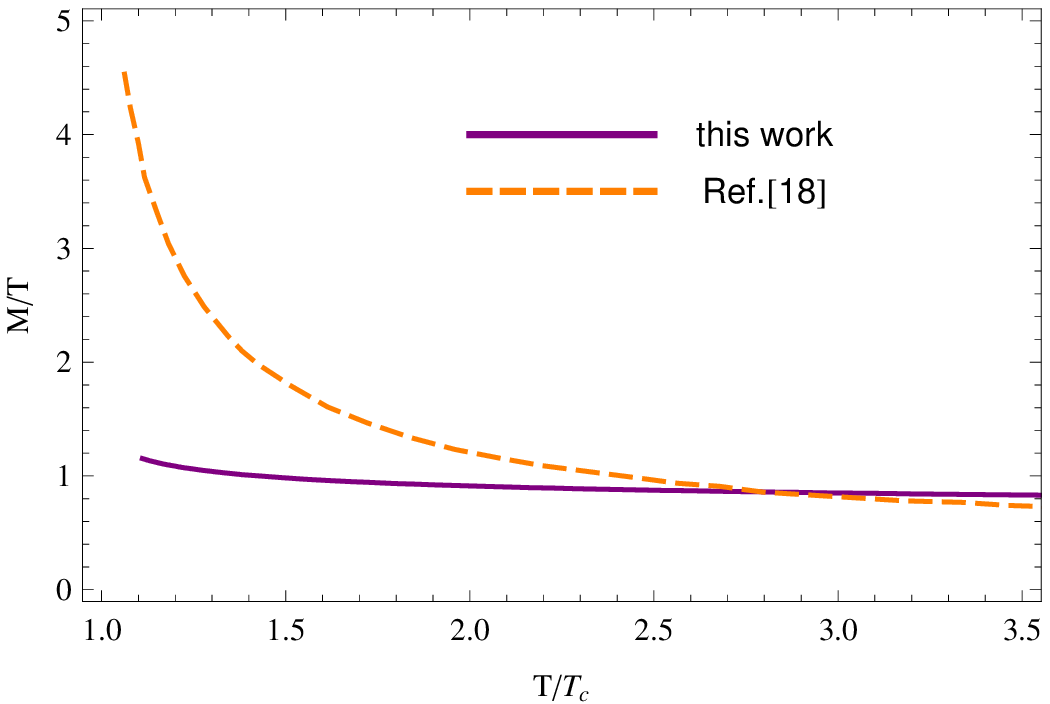}~~~~~
\includegraphics[width=7.5cm]{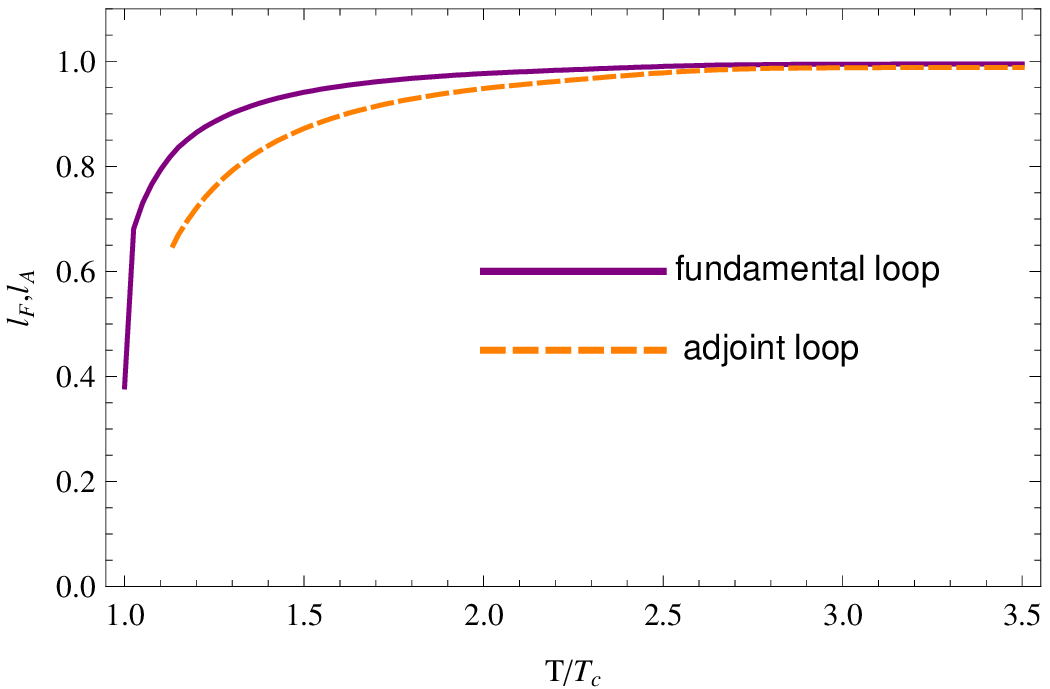}
\caption{\label{Fig:masses} {\em Left panel}. Ratio $M/T$ for our model (indigo, solid line) and 
for the model without Polyakov loop (orange, dashed line) of~\cite{Castorina:2011ra}, against
temperature, the latter measured in units of the critical temperature $T=270$ MeV.
{\em Right panel}. Expectation value of the fundamental (indigo, solid line) and adjoint
(orange, dashed line) loops against temperature.
Figure adapted from Ref.~\cite{Ruggieri:2012ny}.}
\end{figure}

In the left panel of Fig.~\ref{Fig:masses} we plot our result for the gluon mass
against temperature, and we compare it with the result of~\cite{Castorina:2011ra} obtained without the
Polyakov loop dynamics. In the right panel of the same figure we plot our results
for the fundamental and the adjoint loops.

It is interesting that when the Polyakov loop is not introduced, hence within the common
gluon plasma nature of the deconfinement phase, a large mass is needed
to suppress pressure and energy density around $T_c$. As it is discussed in~\cite{Castorina:2011ra}, 
this divergence embeds the nonperturbative effects which are important at the phase transition. 
In our case, the statistical suppression of states below and around $T_c$ is achieved by virtue of the Polyakov loop,
similarly to what happens in the
PNJL model~\cite{Fukushima:2003fw,Ratti:2005jh,Fukushima:2008wg,Abuki:2008nm}. 
For this reason, we do not need to have a large mass as $T$ approaches $T_c$.
The nonperturbative behavior of the Polyakov loops as a function of the temperature is
assured by the combined effect of the Polyakov loop effective potential, 
which is dominant at low temperatures and favors the confinement phase,
and the quasiparticle potential, which is dominant at large temperatures and
favors a nonzero expectation value of the Polyakov loops.
Being the Polyakov loops different from zero, the deconfinement phase within our theoretical description is characterized
by a twofold nature: a sea of background traced $Z(N)-$lines in which transverse
quasiparticles propagate.

\section{Conclusions}
In this talk, we have summarized our recent results~\cite{Ruggieri:2012ny}
about a combined description of the $SU(3)$ deconfinement phase,
in terms of gluon quasiparticles and Polyakov loops. This picture merges 
the common view of the deconfinement phase as a plasma, with a picture
where the relevant degrees of freedom are traced $Z(N)-$lines (Polyakov loops
in fundamental as well as in higher representations)~\cite{Pisarski:2006hz}.
Our main purpose is twofold. Firstly, we are interested to a simple description of
the lattice data about the thermodynamics of the gluon plasma. This is interesting
because it allows to understand which are the relevant degrees of freedom 
in the deconfinement phase of the theory. Moreover, once the proper degrees
of freedom are identified, the effective description studied here can be completed
by adding dynamical quarks.
Employing a relativistic transport theory, this can allow also a direct connection 
between the developments of effective models and the study of the quark-gluon plasma
created in relativistic heavy-ion collisions.
Within our model, we are able to reproduce fairly well the lattice data about
the gluon plasma thermodynamics in the critical region. 

One of the main conclusions of our work is that because of the coupling to the Polyakov loop,
a gluon mass of the order of the critical temperature, $m_g\sim 1-2\, T_c$, is enough to reproduce
the lattice data in the critical region. This is different from what is found
in standard quasi-particle models, which support the pure plasma picture
of the deconfinement phase: in that case the Polyakov loops
are neglected and the phase is described in terms of a perfect gas of
massive gluons. As a consequence, a mass rapidly increasing with $T$
is needed to suppress pressure and energy density in the critical region. 
On the other hand, in the case studied in this article, the suppression of
the pressure and energy density is mainly caused by the Polyakov loop. 
Therefore, lighter quasiparticles with a smooth 
$T-$dependence
describe fairly well the lattice data in the critical region.

There are several interesting directions to follow to extend the present
work. Firstly, having in mind the study of the quark-gluon plasma,
we plan to add dynamical massive quarks to the picture. 
It is also of a certain interest to span the parameter
space in more detail, eventually analyzing several
functional forms of the gluon mass. 
Moreover, even in the case of the pure gauge theory, it would be interesting
to investigate the way to adjust the interaction
measure in order to reproduce lattice data in the regime of
very large temperature. Even more, it would be of a certain interest
to extend our study to the case of different gauge groups.

\begin{theacknowledgments}
The speaker (M. R.) acknowledges the organizers for their warm acceptance
at the workshop venue. Moreover, he acknowledges fruitful discussions during the
workshop with H. Abuki, M. Alford, M. Chernodub, M. Frasca, M. Huang, S. Nicotri, T. Mukherjee
and M. Tachibana. 
This work was supported in part by the Italian Ministry
of Education, Universities and Research under the FIRB Research Grant
RBFR0814TT. 
\end{theacknowledgments}

\bibliographystyle{aipproc}   

\end{document}